\documentclass[openacc]{rstransa}%%%%where rstrans is the template name

\begin{document}

%%%% Article title to be placed here
\title{Physical Characteristics of Jupiter's Trojan (1437) Diomedes from a Tri-chord Stellar Occultation in 2020 and Dimensionless 3D Model}

%%%% Author details
\author{
H. Dutra$^{1}$,
M. Assafin$^{1,2}$,
B. Sicardy$^{3}$,
J. L. Ortiz$^{4}$,
A. R. Gomes-J\'unior$^{5,2}$,
B. E. Morgado$^{1,2}$,
G. Benedetti-Rossi$^{2,6}$,
F. Braga-Ribas$^{7,2}$,
G. Margoti$^{7,2}$,
E. Gradovski$^{7,2}$,
J. I. B. Camargo$^{8,2}$,
R. Boufleur$^{2}$,
R. Vieira-Martins$^{8,2}$,
J. Desmars$^{3}$,
D. Oesper$^{9}$,
K. Bender$^{10}$,
C. Kitting$^{11}$,
R. Nolthenius$^{9,12}$
}

%%%%%%%%% Insert author address here
\address{\small $^{1}$Universidade Federal do Rio de Janeiro - Observat\'orio do Valongo, Rio de Janeiro, Brazil; $^{2}$Laborat\'orio Interinstitucional de e-Astronomia - LIneA, Rio de Janeiro, RJ, Brazil; $^{3}$Observatoire de Paris, France; $^{4}$Instituto de Astrofísica de Andalucía – Consejo Superior de Investigaciones Científicas, Granada, Spain; $^{5}$Universidade Federal de Uberl\^andia (UFU), MG, Brazil; $^{6}$Universidade Estadual de S\~ao Paulo (UNESP), Guaratinguet\'a, SP, Brazil; $^{7}$Federal University of Technology - Paran\'a (PPGFA/UTFPR-Curitiba), Curitiba, PR, Brazil; $^{8}$Observatório Nacional/MCTI, Rio de Janeiro, RJ, Brazil; $^{9}$International Occultation Timing Association (IOTA), Fountain Hills, AZ, USA; $^{10}$Cabrillo College, Aptos, CA, USA; $^{11}$Biological Sciences, California State University, Hayward, CA, USA; $^{12}$ Cabrillo College and Earth Futures Institute, UC Santa Cruz, 6500 Soquel Drive, Aptos, CA 95003, USA.
}

%%%% Subject entries to be placed here %%%%
\subject{Solar System astrophysics, minor planets.}

%%%% Keyword entries to be placed here %%%%

%%%% Insert corresponding author and its email address}
\corres{Hélio Honório Dutra\\
\email{hdutra@ov.ufrj.br}}

%%%% Abstract text to be placed here %%%%%%%%%%%%
\begin{abstract}
\small
Jupiter Trojans preserve primitive formation characteristics due to their collisionless stable orbits. Determination of their shapes and size-frequency distribution constrains the collisional evolution of their parent population which also originated the Kuiper Belt. We started a program to find precise sizes/shapes for Trojans, combining stellar occultations and DAMIT 3D shape models. We report results for Diomedes, by fitting its dimensionless 3D model to 3 chords of a stellar occultation observed in 2020, using iterative $\chi^{2}$ procedures. The pole coordinates, rotation period, volume-equivalent radius and geometric albedo were: $\lambda$ = 153.73$^{o}$ $\pm$ 2.5$^{o}$, $\beta$ = 12.69$^{o}$ $\pm$ 2.6$^{o}$, $P$ = 24.4984 $\pm$ 0.0002 h, $R_{eq}$ = 59.4 $\pm$ 0.3 km and $p_{V}$ = 0.030 $\pm$ 0.004. A precise position was obtained too.
\end{abstract}
%%%%%%%%%%%%%%%%%%%%%%%%%%%

%%%%%%%%%% Insert the texts which can accomdate on firstpage in the tag "fmtext" %%%%%

%\begin{fmtext}

%\end{fmtext}

%%%%%%%%%%%%%%% End of first page %%%%%%%%%%%%%%%%%%%%%

\maketitle

\section{Introduction}\label{sec:intro}

Jupiter's trojan asteroids or Trojans \cite{bottke} are located at the Sun-Jupiter Lagrangian points L4 and L5 at 60$^{o}$ separation angles from the planet, following its orbital movement. They tend to be at these locations indefinitely with rare collision events due to the gravitational stability of these regions \cite{faure}. Because they are not large bodies and due to the relative absence of collisions at L4 and L5, Trojans have undergone few physical changes and, hence, have preserved their primitive characteristics since their formation. Since Trojans came from the same parent population that also sculpted the Kuiper Belt \cite{nesvorny}, the establishment and refinement of the size-frequency distribution (SFD) and shapes of Trojans (among other physical characteristics) contribute not only to the understanding of Trojans themselves, but also to the knowledge of their parent population in the early Solar System. This allows us to restrict models and scenarios of the formation and evolution of the Solar System before and after the dissipation of the solar nebula, including an indication of the most likely dynamic processes shaping our current Solar System \cite{nesvorny}. 

This work presents pilot results from a long term program aiming at the determination of the sizes and shapes of Trojans by the use of stellar occultation techniques combined with dimensionless 3D shape models from the DAMIT database \cite{damit}. By using a 3D model and the 3 positive chords of a stellar occultation observed on November 1st, 2020, we successfully improved the measurements of physical characteristics of Trojan (1437) Diomedes, namely the pole orientation, rotational parameters and sizes of the 3D model, mass and geometric albedo.

Section \ref{sec:obs} summaries the stellar occultation prediction and observations. Section \ref{sec:LCs} describes how the occultation light curves were derived, with the determination of ingress/egress instants given in Section \ref{sec:instants}. The method for fitting 3D shape models to chords projected in the sky plane is presented in Section \ref{sec:3Dfit}, where we describe the application and results for the case of Diomedes. From the fit, we give in Section \ref{sec:characteristics} the derived physical characteristics of Diomedes. Discussions in Section \ref{sec:conclusions} present comparisons with other Diomedes measurements, prospects for oncoming improvements of the method, and the future of our Trojan program.

\section{Stellar occultation: prediction and observations}\label{sec:obs}

\begin{figure}[h!]
    \centering
    \includegraphics[scale=0.26]{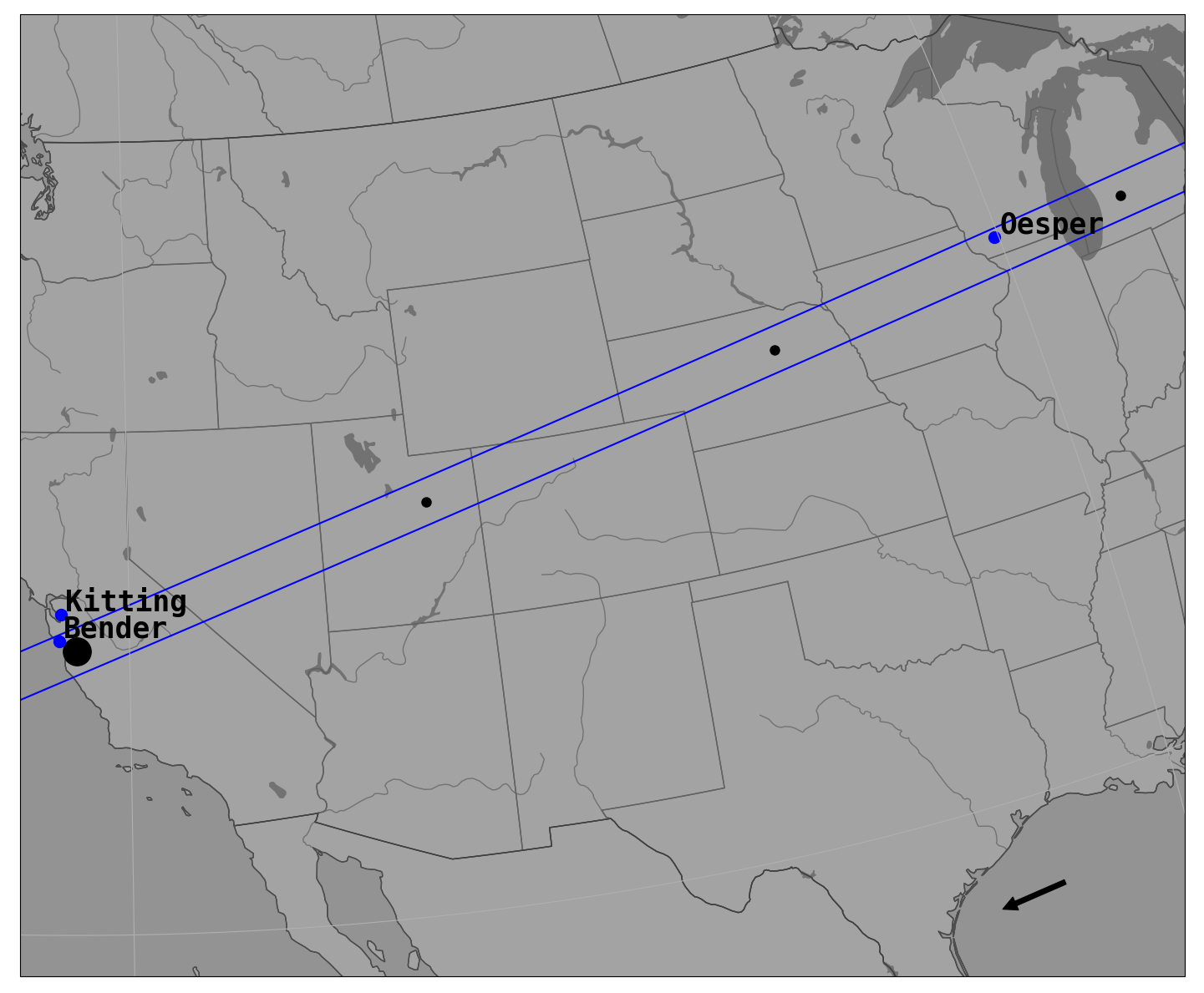}
    \caption{Prediction map of the occultation event by Diomedes on November 1st, 2020 at instant 06:55:54.760 (UTC). The blue lines show the predicted occultation shadow path crossing the Earth. The location of observers resulting in positive chords are indicated by blue dots. The black dots are separated by 60 s time intervals and the largest one indicates the geocentric closest approach of the shadow. The black arrow at the bottom right shows the direction of the shadow.}
    \label{fig:map_zoom}
\end{figure}

The occultation event of the star GAIA DR3 322153921937233152 (mag G $\approx$ 13.59) by Diomedes was predicted by the Lucky Star team. The predicted instant of this event was at 06:55:54.760 (UTC) on November 1st, 2020 and the geocentric velocity of the shadow, which crossed the United States - see figure \ref{fig:map_zoom} - was -16.49 km s$^{-1}$. The separation angle Sun-geocenter-star was 154.79$^{o}$ and Moon-geocenter-star, 29.78$^{o}$. Information about star and Diomedes for this event is shown in Table \ref{tab:starbody}. Three positive detections were reported by three observers located at different regions inside and near the predicted shadow path. All observers are IOTA\footnote{\url{https://occultations.org/}} members. Information about their location and instrumental settings are listed in Table \ref{tab:obs}.

No filters were used in the observations. Time instants were inserted at each field of the acquired AVI video files by using IOTA-VTI\footnote{\url{https://videotimers.com/home.html}} and KIWI-OSD\footnote{\url{https://sites.google.com/site/kiwiosd}} devices equipped with GPS receivers. The AVI files were converted to FITS images \cite{wells1981} by using TANGRA version 3.7.3 software\footnote{\url{http://www.hristopavlov.net/Tangra3/
}}. The time instants for each FITS image were calculated by manipulating the AVI-extracted fields and frames according to the exposure and cycle times, following the procedures described in \cite{guga2016} and \cite{barry2015}.

\begin{table*}[h!]
    \begin{center}
    \scalebox{0.9}{
    \begin{tabular}{ll} % two columns, alignment for each
        \hline
        \hline
    \multicolumn{2}{c}{Occulted star}\\
		\hline
    Epoch                       &  2020-11-01 06:55:54.760 UTC\\
    Source ID                   &  Gaia DR3 322153921937233152\\
    Star position $^{(1)}$ & $\alpha$ = 01$^h$ 32$^m$ 14.$^s$97584 $\pm$ 0.0613 mas\\
                                & $\delta$ = 36$^o$ 36$^\prime$ 37.$^{\prime\prime}$7762 $\pm$ 0.0702 mas\\
    Magnitudes $^{(2)}$         & G = 13.593; B = 13.820; V = 13.360; R = 13.400\\
                                & J = 12.750; H = 12.510\\
    Absolute magnitude G $^{(3)}$ & M$_{G}$ = 2.988	\\
    Apparent diameter $^{(4)}$  & 0.0131 mas = 0.03918 km\\
		\hline
    \multicolumn{2}{c}{Diomedes}\\
		\hline
    Ephemeris    $^{(5)}$         & JPL$\#$88 DIOMEDES/DE440\\
    Geocentric distance          & 4.1323989 AU\\
    Shadow velocity              & -16.49 km s$^{-1}$ \\
    Apparent magnitude             & V = 15.247\\
    Diameter $^{(6)}$              & 117.786 $\pm$ 1.179 km\\
    Geometric albedo (V) $^{(7)}$ & 0.061 $\pm$ 0.011 \\
    Rotational period $^{(8)}$    & 24.4987 $\pm$ 0.0002 h\\
    Absolute magnitude $^{(9)}$    & H$_{V}$ = 8.99 $\pm$ 0.14 \\
    
    	\hline  
    \end{tabular}}
    \end{center}
    \caption{Information about the target star and Diomedes. (1) Star position \cite{gaia2022} computed by SORA \cite{sora} at instant of the occultation event following proper motion corrections \cite{cantat2021}. (2) magnitudes data from NOMAD catalog I/297/out \cite{nomad}. (3) Value from GAIA catalog I/355/paramp \cite{gaia2022}. (4) Apparent diameter of the star at the Diomedes distance for the occultation event instant following the Kervella model \cite{kervella2004}. (5) Data from JPL Horizons e JPL NAIF web service. (6) and (7) \cite{mainzer2019}. (8) \cite{durech2019}. (9) See Section \ref{sec:characteristics}.}
%    \end{center}
    \label{tab:starbody}
\end{table*}

\begin{table}[h!]
    \begin{center}
    \scalebox{0.9}{
        \begin{tabular}{ccccc}
        \hline
        \hline
        \multicolumn{1}{c}{Observer} & \multicolumn{1}{c}{Latitude N ($^o ~ ^\prime ~ ^{\prime\prime}$);} & \multicolumn{1}{c}{D(cm); f/;} & \multicolumn{1}{c}{\underline{Observation (UTC)}} & \multicolumn{1}{c}{Exp. time (s);} \\ 
        
        & Longitude E ($^o ~ ^\prime ~ ^{\prime\prime}$); & Detector; & Start (06:mm:ss); & Cycle time (s); \\       
        
        &  Altitude (m) & Format & End (06:mm:ss) & Time device \\
        \hline 
        
        Chris Kitting, USA & 37 38 48.84;    & 25.40; 05.0;  & 04:30     &  0.2670; \\
                       & -122 02 09.096; & Watec 910HX;  & 56:27.13  &  0.2670; \\
                       & 620             & AVI           &           &  IOTA-VTI \\ \\
        Kirk Bender, USA   & 37 01 19.82;    & 20.32; 05.0;  & 54:48.226 &  0.2670; \\
                       & -122 05 07.09;  & Watec 910HX;  & 57:53.725 &  0.2670; \\
                       & 358             & AVI           &           &  IOTA-VTI \\ \\
        David Oesper, USA  & 42 57 36.9;     & 30.50; 03.3;  & 43:00     &  0.1330; \\
                       & -90 08 31.1;    & Watec 910HX;  & 60:00     &  0.1331; \\
                       & 390             & AVI           &           &  KIWI-OSD \\ 
        \hline
        \end{tabular}
        }
        \end{center}
        \caption{Location of the observers and their respective observational information. D(cm)= diameter. f/ = focal ratio. The observation (UTC) time refers to November 1st, 2020.}
    \label{tab:obs}
\end{table}

\section{Stellar occultation light curves}\label{sec:LCs}

\begin{figure}[h!]
    \centering
    \includegraphics[scale=1]{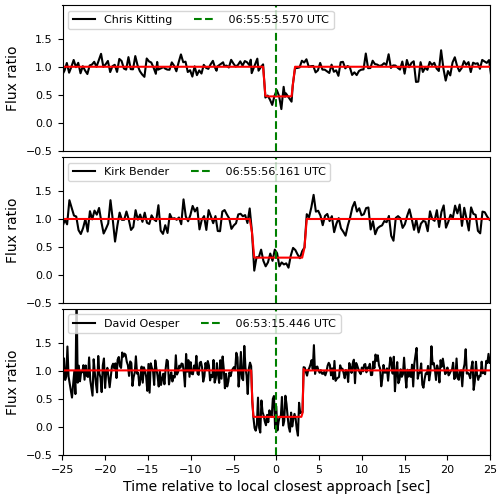}
    \caption{Best fitted light curves (in red) for the observed light curves (black dots and lines) for each observer. The observed flux ratios were normalized outside the events. The green vertical lines indicate the mid-instants of the occultation from the fits for each observer.}
    \label{fig:lc}
\end{figure}

Differential aperture photometry over the FITS images was carried out using the PRAIA (Package for the Reduction of Astronomical Images Automatically) package \cite{assafin2023}. This package is a suitable tool for data treatment of stellar occultations. The PRAIA Photometric Task (PPT) selects comparison stars to normalize the flux ratio (target/calibrators) in order to generate a light curve with the least dispersion ($\sigma$). All light curves are plotted in figure \ref{fig:lc}. The flux drops in all curves (black lines) are near the predicted instants for each observer (measured mid-instants are indicated by green lines). The fitted light curves (in red) will be described in the next section.

In Figure \ref{fig:lc} it is possible to notice that the light curve minimum at the occultation instants is never zero, being smaller for Oesper and Bender than Kitting's observations. As can be seen in Table \ref{tab:obs}, although all observers use the same detector, the telescope aperture, exposure time and the sky transparency will determine the signal-to-noise ratio, i.e. how much of the residual reflected light from the body will still be detected during occultation. The first two observers from Table \ref{tab:obs} set the same exposure time, but the Kitting's telescope aperture is larger. On the other hand, even though the Oesper's telescope aperture is the largest one among all observers, the exposure time was set to about half the value used by the others.

\section{Ingress and egress occultation instants}\label{sec:instants}

The ingress and egress instants of each light curve were determined by an iterative $\chi^{2}$ procedure resorting to the SORA (Stellar Occultation Reduction Analysis) package \cite{sora}. Considering an opaque body without atmosphere, instants are defined by a geometric sharp-edge light curve occultation model (box model), convolved with effects caused by the apparent star diameter at the body's geocentric distance, Fresnel diffraction, shadow velocity, exposure time and dead time (readout time). The box model is restrained by the normalized flux ratios outside of the event (unity) and inside it (bottom of the observed light curve). Simulated light curves are generated by spanning different ingress/egress instants. For each simulation, the $\chi^{2}$ with respect to the observed light curve is computed. The measured ingress/egress instants result from the best fit (simulated light curve with the least $\chi^{2}$, or $\chi^{2}_{min}$). Instant error values come from considering ingress/egress instants resulting in $\chi^{2}$ values equal to $\chi^{2}_{min}$ $+$1 and $\chi^{2}_{min}$ $+$9 (1 and 3$\sigma$ errors, respectively). 

\begin{figure}[h!]
    \centering
    \includegraphics[scale=0.58]{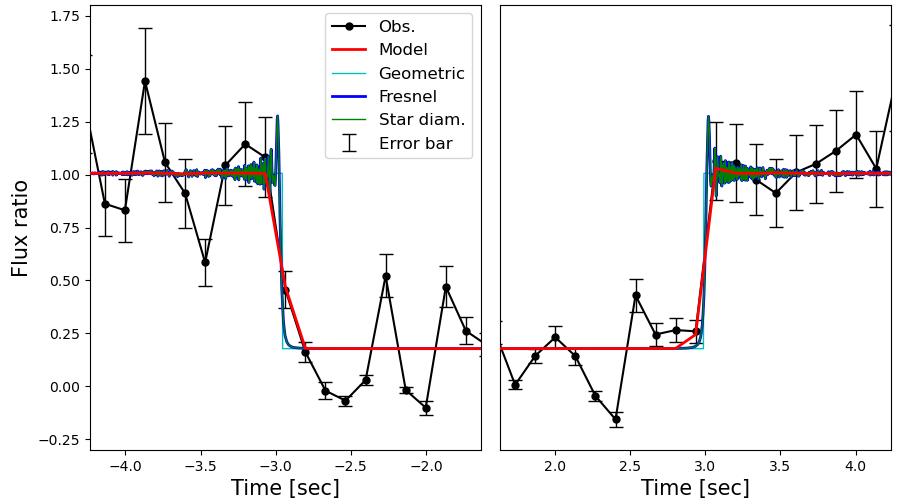}
    \caption{Light curve fitting for D. Oesper's data. The black dots/lines are the observed light curve obtained with the PRAIA package, with 1 $\sigma$ error bars displayed. The best fitted light curve in red comes from an interactive $\chi^{2}$ procedure using the SORA package (see text), by the convolution of a sharp-edge geometric (box) model (light blue curve) with effects caused by the Fresnel diffraction (blue curve), apparent star diameter (green curve), and local normal star velocity coupled with the exposure and dead (readout) times. The obtained ingress/egress instants and errors come from the associated box model. The normalized flux ratio is plotted as a function of the seconds before and after the local closest approach (06:53:15.446 UTC).}
    \label{fig:lc_zoom}
\end{figure}

The occultation star has 4.934 solar luminosity, effective temperature equals 6483.5680 K and its spectral type is F6 \cite{gaia2022}, i.e. it is a main sequence dwarf in the H-R diagram. Using the Kervella model \cite{kervella2004} and van Belle model \cite{Belle1999} to calculate the star diameter based on its apparent magnitudes (see the Table \ref{tab:starbody}), we get 0.0131 mas and 0.0122 mas, respectively, in V band. Therefore, the star diameter at the geocentric distance (D) of Diomedes at the occultation instant (see Table \ref{tab:starbody}) is $\approx$ 0.04 km using both models. Assuming a typical wavelength $\lambda$ = 0.7 $\mu$m, the Fresnel scale at the distance D is $L_{f} = \sqrt{\lambda D 2^{-1}}$ = 0.46 km. The shortest exposure time used was 0.133 s (see Table \ref{tab:obs}) or 2.19 km given the shadow velocity (see Table \ref{tab:starbody}). Therefore, the exposure time is the dominant effect over the light curves. Figure \ref{fig:lc_zoom} illustrates the light curve fitting of Oesper's data as example. The best light curves fitted for each observer were displayed in the figure \ref{fig:lc}. Table \ref{tab:chords} displays the fitted ingress and egress instants for each observer, the corresponding chord lengths in the sky plane and the standard deviation of the flux ratio computed outside of the occultation. All instants displayed in Table \ref{tab:chords} were corrected by systematic time delays \cite{barry2015} from the instrument and camera configuration setup used by the observers. Thus, a time offset of -0.284 s was applied on the ingress and egress instants from K. Bender and C. Kitting, and -0.150 s for D. Oesper instants, by following tables of video time insertion analysis available on website by occultation community\footnote{\url{http://www.dangl.at/ausruest/vid\textunderscore tim/vid\textunderscore tim1.htm}}. Those values were updated in figure \ref{fig:lc}.

\begin{table*}[h!]
    \begin{center}
    \scalebox{0.9}{
    \begin{tabular}{ccccccc}
    \hline
    \hline
    Observer  & \multicolumn{2}{c}{Ingress (UTC)} & \multicolumn{2}{c}{Egress (UTC)}   & Chord length & $\sigma$ \\ 
                & (06:mm:ss.s) & 1 $\sigma$(s) & (06:mm:ss.s) & 1 $\sigma$(s)  &      (km)      & (flux) \\
                &              &1 $\sigma$(km) &              & 1 $\sigma$(km) & & \\ \hline
    C. Kitting  &   55:51.888  & 0.026         &  55:55.252   &   0.033        & 55.459 $\pm$ 0.973 & 0.086 \\
                &              & 0.429         &              &   0.544        & & \\
    K. Bender   &   55:53.084  & 0.058         &  55:59.241   &   0.056        & 101.504 $\pm$ 1.879 & 0.146 \\
                &              & 0.956         &              &   0.923        & & \\
    D. Oesper   &   53:12.474  & 0.023         &  53:18.419   &   0.023        & 98.009 $\pm$ 0.759 & 0.149 \\
                &              & 0.379         &              &   0.379        & & \\ \hline
    \end{tabular}}
    \end{center}
    \caption{Fitted ingress and egress instants (and 1 $\sigma$ errors) for each observer, and respective chord lengths. The $\sigma$ (flux) is the observed normalized light curve flux ratio dispersion computed outside of the occultation instants.}
    \label{tab:chords}
\end{table*}

\section{3D model limb fitting of Diomedes}\label{sec:3Dfit}

SORA provides the necessary features to fit the limb of three-dimensional body models (OBJ file format) to occultation chords projected in the sky plane, given input values for the rotational velocity, pole coordinates (in ICRS\footnote{International Celestial Reference System}) and precession parameters. In this way, the body's dimensions, rotation and orientation parameters can be improved.

A 3D shape model of a body is represented as a polyhedron with triangular surface facets. The size or scale of the model is dictated by the distances of the vertices of the triangles to the center. Vertex distances can be dimensionless or may be expressed in some unity, e.g. in km. Multiplying vertex distances by a scale factor alters the size of the model. If the 3D model is dimensionless, we can multiply its vertex distances by a scale factor in km to resize it in km.

The aspect of the body's limb as projected in the Bessel plane (roughly the sky plane) is governed by the \textit{scale} of the 3D model, the projected $(f, g)$ center, and by the space orientation of the body, described by its sub-observer elements in the planet-centric system: the sub-observer longitude ($Lon$), sub-observer latitude ($Lat$) and pole's position angle ($PA$). Sampling these 6 parameters allows us to find the best limb match with the extremities of the projected occultation chords in an iterative $\chi^{2}$ procedure in terms of radial limb offsets. The best parameter value results in $\chi^{2}_{min}$, with 1 and 3 $\sigma$ errors estimated by the parameter's $\chi^{2}$ distribution obtained from sample runs with $\chi^{2}$ values of $\chi^{2}_{min}$ $+$1 and $\chi^{2}_{min}$ $+$9 respectively. The fitted sub-observer parameters are then converted back to pole orientation parameters in the usual equatorial (Archinal) or ecliptic (Kaasalainen) systems \cite{archinal} \cite{kaasalainen} in an iterative procedure. The comparison between the expected (nominal) and fitted sub-observer longitude ($Lon$) allows for improving the rotation period. Offsets between the predicted (0,0) center and the fitted $(f, g)$ one, together with the star position, allow to derive a usually precise astrometric position. This method was applied for Trojan Diomedes, as described next.

Diomedes has a convex 3D model composed of 574 vertices and 1144 facets \cite{durech2019} derived by light curve inversion techniques \cite{kaasalainen}, i.e. resulting in dimensionless vertex distances arbitrarily scaled to give the model a unit volume. The model can be found in DAMIT\footnote{\url{https://astro.troja.mff.cuni.cz/projects/damit/} (updated webpage address)}(Database of Asteroid Models from Inversion Techniques) database \cite{damit}. Its rotational parameters are also given in LCDB\footnote{\url{https://alcdef.org/}} (Light Curve Data Base). This 3D model was constructed with a large photometric data provided by the Lowell Observatory and GAIA DR2. The coverage period from Lowell Observatory was October 1998 -- March 2012, containing 320 observations spread along Diomedes' orbit around the Sun, with typical precision of 0.1 -- 0.2 magnitudes in V band. GAIA dataset covers August 2014 -- May 2016 with 26 sparse observations with precision around 0.01 magnitudes in G band \cite{durech2019}. The provided pole's longitude and latitude in ecliptic system are respectively $\lambda$ = (150 $\pm$ 5)$^{o}$ and $\beta$ = (5 $\pm$ 5)$^{o}$. Its rotational period is P = 24.4987 $\pm$ 0.0002 h. The initial rotation angle is $\phi$ = 0$^{o}$ for the reference epoch JD0 = 2451104 (October 17th, 1998 at 12:00:00 UTC).

\begin{figure}[h!]
    \centering
    \hspace*{-1.2cm}
    \includegraphics[scale=0.58]{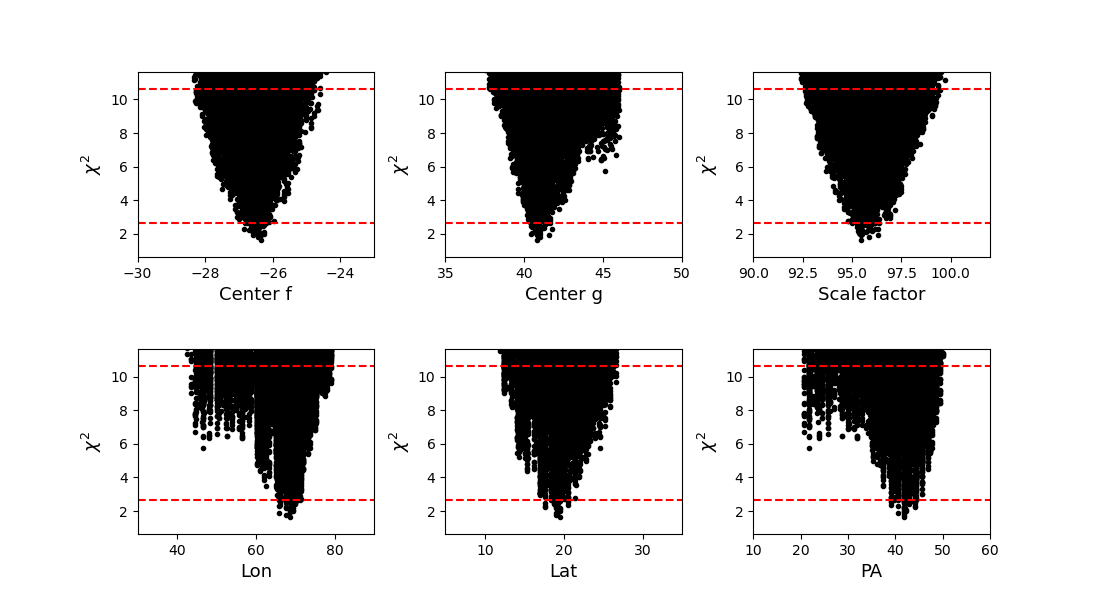}
    \caption{$\chi^{2}$ distributions for sub-observer parameters ($Lon$, $Lat$, $PA$), $(f, g)$ center and \textit{scale factor} after many samplings in the fitting of the 3D Diomedes model to the 3 occultation chords of the 2020 stellar occultation. Red lines indicate $\chi^{2}$ values of $\chi^{2}_{min}$ $+$1 and $\chi^{2}_{min}$ $+$9, associated with the computation of respectively 1 and 3 $\sigma$ errors for the parameters.}
    \label{fig:chi2}
\end{figure}

The fitting procedure for Diomedes was as follows. For each set of input values for parameters $(f, g)$, $Lon$, $Lat$, $PA$ and \textit{scale}, SORA updates the 3D model aspect in space and computes the corresponding limb in the Bessel plane for the geocentric reference at the 2020 occultation instant. By \textit{scale}, we mean a scale factor which multiplies the original vertex distances to size the model in km. Also, ecliptic pole coordinates are converted to the equatorial system used by SORA, to obtain the prime angle equation ($W$) and, finally, to get the corresponding sub-observer longitude ($Lon$), sub-observer latitude ($Lat$) and position angle ($PA$). For each set of input values for the parameters, SORA computed radial offsets between the limb and the extremities of each of the 3 occultation chords, and stored the resulting $\chi^{2}$. Since the orientation parameters ($Lon$, $Lat$, $PA$) should be reasonably consistent with the nominal pole orientation and rotation period implicit in the Diomedes 3D model \cite{durech2019}, we did not need to test all possible values, avoiding high computational costs. However, due to the limited number of data points (chord extremities), a fair range of values were allowed for each parameter, while still taking into account the uncertainties associated to the nominal rotation period, pole coordinates and ephemeris of Diomedes. For the same reason, a huge number of samplings was done (528 million runs). The obtained $\chi^{2}$ distributions for each parameter are shown in figure \ref{fig:chi2}. From these distributions, the fitted parameters and errors are obtained. Table \ref{tab:3d_fit} lists nominal values associated to the 3D Diomedes model and the fitted values obtained for each parameter.

\begin{table}[h!]
    \centering
    \scalebox{0.9}{
    \begin{tabular}{c|c|c}
    \hline \hline
        Parameters & Nominal values & Fitted values \\
        \hline 
        Center f & 0 & -26.551 $\pm$ 0.300 km \\ 
        Center g & 0 & 41.094 $\pm$ 0.649 km \\ 
        Sub-observer longitude $Lon$ & 103.611$^{o}$ & 67.998 $\pm$ 2.063$^{o}$ \\ 
        Sub-observer latitude $Lat$ & 19.787$^{o}$ & 19.078 $\pm$ 1.482$^{o}$\\ 
        Position angle $PA$ & 50.545$^{o}$ & 41.587 $\pm$ 2.491$^{o}$\\ 
        Scale factor &  1 & 95.816 $\pm$ 0.552 km\\ \hline
    \end{tabular}}
    \caption{Nominal values of the sub-observer parameters ($Lon$, $Lat$, $PA$), $(f, g)$ center and \textit{scale factor} at the instant of occultation and the results obtained by fitting these parameters with their 1 $\sigma$ errors.}
    \label{tab:3d_fit}
\end{table}

\begin{figure}[h!]
    \centering
    \includegraphics[scale=0.35]{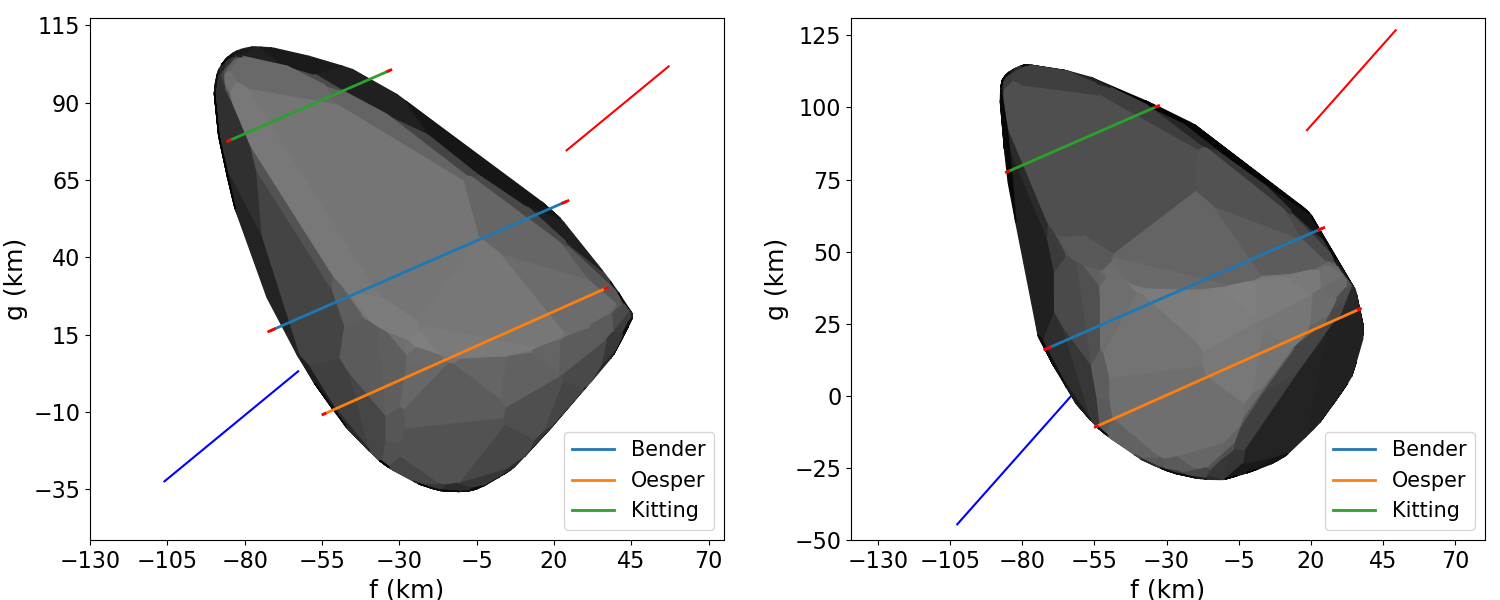}
    \caption{3D model of Diomedes using six parameters ($f$, $g$, scale factor, sub-observer longitude $Lon$ and latitude $Lat$, pole's position angle $PA$) projected in the Bessel plane with the chords at the instant of the occultation event. Left panel: projection of the 3D model with the nominal values of the four parameters except for $f$ and $g$, which were fitted to the chords. Right panel: projection with the fitted values. The same (fitted) scale factor was used in both plots. Detached orange (north) and blue (south) lines indicate the rotation axis. The improvement on Diomedes' pole orientation and rotational parameters are clear.}
    \label{fig:3d_model}
\end{figure}

Figure \ref{fig:3d_model} illustrates the situation before and after the fitting process. The projections of the 3D model in the sky plane uses the nominal and fitted values listed in Table \ref{tab:3d_fit} with the same (fitted) scale factor. It is clear that the fit improved Diomedes' pole orientation and rotational parameters.

Since the Gaia DR3 catalogue position errors of the occulted star are quite small (Table \ref{tab:starbody}), the fitted $(f, g)$ center directly furnishes precise body’s ephemeris offsets with respect to its predicted astrometric position represented in position (0, 0) in figure \ref{fig:3d_model}. With these offsets and the Gaia DR3 star position, we obtain from the 3D model fit a precise astrometric position at the instant of the occultation, as listed in Table \ref{tab:position2}.

\begin{table}[h!]
    \centering
    \scalebox{0.9}{
    \begin{tabular}{c|c|c}
    \hline \hline
    Right ascension (h m s)   & Astrometric uncertainty & ($\Delta \alpha cos\delta$, $\Delta \delta$) ephemeris offsets \\
    Declination ($^{o}$ ’ ”)  &   (mas)                 & (mas)              \\ \hline
    1 32 14.9752703           &   $\pm$ 0.117           & -8.859 $\pm$ 0.100 \\
    36 36 37.785170           &   $\pm$ 0.228           & +13.711 $\pm$ 0.200 \\
    \hline
    \end{tabular}}
    \caption{Astrometric Diomedes position on 2020-11-01 at 06:55:54.760 (UTC) for the geocentric reference using the 3D model fit. Position offsets regard to the JPL$\#$88 Diomedes/DE440 ephemeris.}
    \label{tab:position2}
\end{table}

\section{Diomedes physical characteristics: rotation period, pole orientation, size, mass, surface gravity, escape velocity, geometric albedo}\label{sec:characteristics}

Given the uncertainty of the nominal rotational period, we should expect a difference of $\pm$ 23.18$^{o}$ in the value of the sub-observer longitude ($Lon$) at occultation instant for an error propagation of 1.5774 h. However, according to Table \ref{tab:3d_fit}, the difference between the nominal and fitted values of $Lon$ was larger than expected: -35.613$^{o}$. Since this difference is equal to the difference in prime angle ($W$) from the equatorial system \cite{archinal}, we can improve the rotation period by using the prime angle equation ($W=W_{o}+\Dot{W}d$), where $W_{o}$ is the initial rotation angle in equatorial system, $\Dot{W}$ is the rotation velocity (360$^{o}$/P(days)) and $d$ is the time interval in Julian Date between the initial reference epoch and the occultation instant. Assuming $\phi_{0}$ and $W_{o}$ as fixed, and using the difference found in $Lon$, the new calculated sidereal period is 24.498393 $\pm$ 0.000006 h. However, considering the photometric noise from the light curves responsible for generating the original 3D model and, as consequence, it's associated nominal uncertainty of 0.0002h on the rotational period, we must combine this uncertainty with that computed. In this way, the new sidereal period becomes P$^{'}$ = 24.4984 $\pm$ 0.0002 h.

To convert sub-observer planet-centric elements and uncertainties to the standard ecliptic system, we randomly sampled 3 million values around the nominal (input) pole coordinates in the ecliptic system within 3 times their formal uncertainties ($\pm$ 15$^{o}$). The corresponding sub-observer planet-centric elements are computed, and we only store those valid pole coordinates whose generated element values lie within the 1 $\sigma$ fitted sub-observer element errors, so that we are considering only 1 $\sigma$ error values for the determination of pole coordinates. In computations, the fitted rotation period was used and the initial phase angle ($\phi_{0}$) was fixed. From the valid values, the obtained pole coordinates in the ecliptic system were $\lambda$ = (153.73 $\pm$ 2.5)$^{o}$ and $\beta$ = (12.69 $\pm$ 2.6)$^{o}$, with 1 $\sigma$ errors taken from the minimum/maximum accepted values.

Table \ref{tab:improvement} displays the nominal and fitted values for the pole orientation and rotation period of Diomedes.

\begin{table}[h!]
    \centering
    \scalebox{0.9}{
    \begin{tabular}{c|c|c}
    \hline \hline
        Parameters & Nominal values & New values \\
        \hline 
        Period & 24.4987 $\pm$ 0.0002 h & 24.4984 $\pm$ 0.0002 h \\
        $\lambda$ & (150 $\pm$ 5)$^{o}$ & (153.73 $\pm$ 2.5)$^{o}$ \\ 
        $\beta$ & (5 $\pm$ 5)$^{o}$ & (12.69 $\pm$ 2.6)$^{o}$ \\ 
         \hline
    \end{tabular}}
    \caption{Nominal and fitted values for the pole orientation and rotation period of Diomedes.}
    \label{tab:improvement}
\end{table}

DAMIT's once dimensionless volume-normalized 3D model for Diomedes is now scaled by a factor equal to 95.816 $\pm$ 0.552 km (Table \ref{tab:3d_fit}), i.e. all vertex distances to the center are now multiplied by this scale factor and have dimension in km. The mean value and standard deviation of these vertex distances are 75.62 (14.17) km, in accord to the irregular shape of the 3D Diomedes model. Considering the scaled model, the total volume is 879658.5 $\pm$ 15115.8 km$^3$, the total surface area is 49559.8 $\pm$ 631.3 km$^2$, and the area delimited by the limb on the Bessel plane at occultation instant is 12154.3 $\pm$ 156.3 km$^2$. For spherical or circular shapes, the corresponding equivalent radius R$_{eq}$ from total volume, surface and limb areas are respectively 59.4 $\pm$ 0.3 km, 62.8 $\pm$ 0.4 km and 62.2 $\pm$ 0.4 km. Table \ref{tab:radius} summaries these results.

\begin{table}[h!]
    \centering
    \scalebox{0.9}{
    \begin{tabular}{c|c|c}
    \hline \hline
    Volume (km$^3$)           & Surface area (km$^2$)   & Limb area (km$^2$) \\
    Equivalent radius (km)    & Equivalent radius (km)  & Equivalent radius (km) \\ \hline
    879658.5 $\pm$ 15115.8    & 49559.8 $\pm$ 631.3     & 12154.3 $\pm$ 156.3 \\
    59.4 $\pm$ 0.3            &  62.8 $\pm$ 0.4         & 62.2 $\pm$ 0.4 \\
    \hline
    \end{tabular}}
    \caption{From DAMIT's now scaled 3D model of Diomedes, the total volume, surface area and limb area at occultation instant, and respective equivalent radii for corresponding spherical or circular shapes.}
    \label{tab:radius}
\end{table}

\begin{figure}[h!]
    \centering
    \includegraphics[scale=0.57]{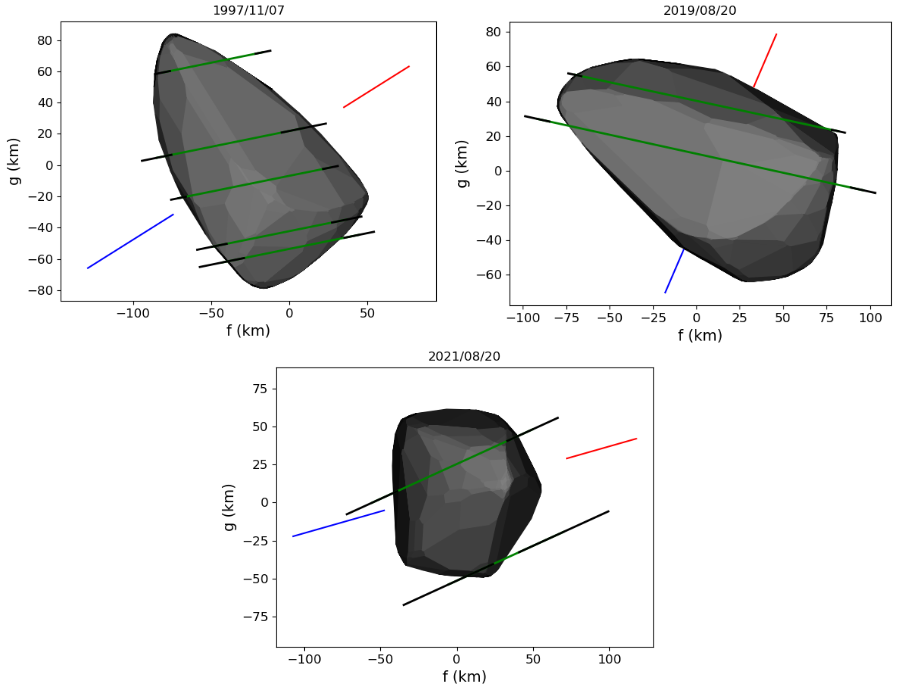}
    \caption{Diomedes new 3D model projected for three more occultation events using the new fitted parameters. Error bars are 3 $\sigma$ values. Here, only $f$ and $g$ parameters were fitted for the chords of each individual event (see text).}
    \label{fig:all_events}
\end{figure}

Three more occultation events occurred on November 7th, 1997, on August 20th, 2019 and August 20th, 2021. The instants of ingress and egress of each event can be found through the Occult software\footnote{http://www.lunar-occultations.com/iota/occult4.htm}. Using these instants, we projected the chords in the Bessel plane as well as the projection of the new Diomedes 3D model according to each occultation epoch, using the fitted scale factor and the new values described in Table \ref{tab:improvement}, except the $f$ and $g$, the only parameters fitted to the chords for each event separately. As can be seen in the Figure \ref{fig:all_events}, the 3D model projections are reasonably consistent with the chords adopting 3 $\sigma$ results for instants of ingress and egress, but there is still room for improvements. Procedures for fitting all parameters with all events are ongoing, but are outside the scope of this work.

In the Tholen classification, Diomedes is a DP type asteroid \cite{neese}, so according to \cite{Krasinsky}, we can infer a density of 1.38 $\pm$ 0.02 g cm$^{-3}$ for Diomedes. This results in a total mass of (1,213,928,730 $\pm$ 38,452,974) 10$^9$ kg for a homogeneous body. From the volume-equivalent radius and this mass, we have an equivalent surface gravity of 2.30 $\pm$ 0.05 cm s$^{-2}$ and equivalent escape velocity of 52.2 $\pm$ 0.7 m s$^{-1}$. We notice however that according to the literature, all these values are only very rough estimates.

If the size/shape and absolute magnitude in V band of the body are known for the occultation instant, its geometric albedo in V band $p_{V}$ can be calculated using equation \ref{eq:albedo} \cite{sicardy2011}, where H$_{\odot,V}$ = -26.74 is the Sun's absolute magnitude in V band, $H_{V}$ is the instantaneous absolute magnitude V of Diomedes (see Table \ref{tab:starbody}), and $R_{eq}$ is the limb-area equivalent radius in km (Table \ref{tab:radius}). $H_{V}$ can be computed for the occultation instant by using a magnitude model (HG, HG$_1$G$_2$ or sHG$_1$G$_2$ models) following the procedures described in \cite{carry}. The absolute magnitude shall be further corrected for rotational phase. We tested all magnitude models with the available 459 photometric measurements from the Zwicky Transient Survey (ZTF) found by the FINK Broker\footnote{\url{https://fink-broker.org/}} (see \cite{carry} and references therein). We got unrealistic values for G$_1$ (almost zero) and absolute magnitude error (1.21) with the HG$_1$G$_2$ model. The new sHG$_1$G$_2$ model incorporates a shape model to fit the data, described by the pole orientation (aspect angle) and oblateness. But perhaps due to incompatibilities with the true 3D shape of Diomedes, the error in absolute magnitude was 1.5, still unacceptably large. The best estimate for the instantaneous absolute magnitude was $H_{V}$ = 8.99 $\pm$ 0.14 with the HG model. No rotation phase correction was applied. We could not derive or find a satisfactory rotational light curve for Diomedes near the occultation epoch. The main difficulty is that its rotation period is approximately 24 hs. But in ALCDEF\footnote{\url{https://alcdef.org/php/alcdef_GenerateALCDEFPage.php}}, one finds a partial rotational curve which clearly displays the minimum and maximum. It was obtained around 1st November 2008 with geocentric solar phase angle $\theta$ = 5.2$^o$ \cite{stephens}, close to the value at the occultation instant, $\theta$ = 4.8$^o$. The amplitude of this light curve is roughly 0.37, which is commensurable with the error found for $H_{V}$. Therefore, a phase correction for absolute magnitude would not make sense. The attained precision for $H_{V}$ takes into account the total error budget. The value obtained for the V band geometric albedo $p_{V}$ was 0.030 $\pm$ 0.004.

\begin{equation}
    p_{V} = \left (\dfrac{149597870.7}{R_{eq}} \right )^{2} 10^{0.4(H_{\odot,V}-H_{V})} 
    \label{eq:albedo}
\end{equation}

\section{Discussion}\label{sec:conclusions}

Using SORA functionalities, we developed a method to fit the limb of 3D body models to occultation chords projected in the sky plane (see Section \ref{sec:3Dfit}). We successfully applied the method to fit the limb of the previously dimensionless 3D model of Diomedes from DAMIT to the three chords of a stellar occultation observed in 2020, turning it into a true 3D model with dimensions in km.

A precise astrometric position is obtained using this method (see Table \ref{tab:position2}). The rotation period and pole orientation of the body are also improved (see Table \ref{tab:improvement}).
The scaled 3D model allows for the determination of the total volume in km$^3$ and surface area in km$^2$, and their corresponding equivalent radii (Table \ref{tab:radius}). Once the density is known, the total mass can be derived, as the surface gravity and escape velocity (Section \ref{sec:characteristics}). After computing the instantaneous absolute magnitude at the occultation instant and applying a rotational phase correction to it, the scaled 3D model also allows for the estimation of the geometric albedo (Section \ref{sec:characteristics}).

The nominal rotational period of Diomedes was 24.4987 $\pm$ 0.0002 h \cite{durech2019}. The expected uncertainty for the occultation instant was thus 23.18$^{o}$ in sub-observer longitude, however a 54\% larger discrepancy of 35.62$^{o}$ was found. Using this difference, we could calculate a new and more precise rotational period for Diomedes (Table \ref{tab:improvement}).

A sampling procedure using the equations relating the pole coordinates in the ecliptic system and sub-observer planet-centric elements was applied, taking into account the respective error ranges for filtering valid values, resulting in the determination of pole coordinates and errors from the fitted sub-observer planet-centric elements. The improved ecliptic pole coordinates found for Diomedes (Table \ref{tab:improvement}) are inside the error bars when compared with literature values: $\lambda$ = (150 $\pm$ 5)$^{o}$ and $\beta$ = (5 $\pm$ 5)$^{o}$ \cite{durech2019}. Our fitted latitude $\beta^{'}$ is closer to one of the solutions found in the work by Hanus et al. (2023), which was 13.42$^{o}$ \cite{hanus}. Notice that our error bars are half the values reported by other authors.

\cite{sato} reported the results of a stellar occultation by Diomedes observed in November 7th, 1997 in Japan around 09$^h$ 29$^m$ 55$^s$ (UTC). The apparent ellipse in the sky plane fitted to the chords had 180 $\pm$ 28 km and 96 $\pm$ 5 km major/minor axes, corresponding to an apparent equivalent radius of 65.7 $\pm$ 6.8 km. Bringing our Diomedes scaled 3D DAMIT model to that epoch results in a limb-area equivalent radius of 64.7 $\pm$ 0.4 km, in excellent agreement with the Japanese measurements.

The equivalent radius of Diomedes measured with the infrared satellite NEOWISE was 58.9 $\pm$ 0.6 km for the mean epoch 2010-02-16 01:22:08.472 UTC \cite{mainzer2019}. Our scaled 3D DAMIT model gives a larger limb-area equivalent radius of 61.8 $\pm$ 0.4 km for this epoch, resulting on a 4.0$\sigma$ difference. The AKARI infrared satellite survey \cite{usui} gives an average equivalent radius of 86.3 $\pm$ 1.7 km from basically 3 distinct epochs (UTC): 2006-05-16 18:50:34.035, 2006-05-17 03:05:45.493 and 2006-11-10 13:35:40.193. Our scaled 3D DAMIT model gives an average limb-area equivalent radius of 67.9 $\pm$ 0.4 km for these epochs, in clear disagreement with these infrared measurements. The IRAS satellite average equivalent radius was 82.2 $\pm$ 2.1 km \cite{iras} from 6 epochs (UTC): 1983-05-26 (20:07:59, 18:24:46), 1983-05-31 13:26:05, 1983-11-09 03:13:52, and 1983-11-15 (12:03:12, 13:46:17). From our scaled model, we get an average limb-area equivalent radius of 67.0 $\pm$ 0.4 km for these epochs, also in disagreement with the infrared result. The V band geometric albedo from NEOWISE \cite{mainzer2019}, AKARI \cite{usui} and IRAS \cite{iras} satellites are respectively 0.061 $\pm$ 0.011, 0.028 $\pm$ 0.001 and 0.031 $\pm$ 0.002. Our obtained geometric albedo was $p_{V}$ = 0.030 $\pm$ 0.004, in full agreement with AKARI and IRAS measurements, and differing within 3$\sigma$ from NEOWISE. In all, the NEOWISE size estimate for Diomedes is closest to the value inferred by the 2020 occultation, but the geometric albedos derived by AKARI and IRAS are more consistent with the occultation value. This suggests there may be a systematic difference in the way that NEOWISE data are being reduced/interpreted compared to the other infrared satellites. However, we don’t have enough asteroid samples analyzed yet using the method discussed in this paper to propose further discussions on this topic.

Fitting the limb of 3D body models to occultation chords allows for the combination of the best characteristics of both techniques: the more precise representation of complex, convex shapes from 3D models, and the more precise determination of sizes, centering, rotation period and pole orientation from occultation chords. The method is certainly useful for breaking the degeneracy in 3D models with two possible pole solutions. In this work, we didn't evaluate the consistency of the fitted parameters with the photometric data from which the 3D model was originally constructed, since the new parameter values are within the nominal error bars of both old and new models, as can be seen in Table \ref{tab:improvement}.

\cite{hanus} presented a scaled 3D model for Diomedes. The shape, pole orientation and rotation period of their 3D dimensionless model come from light curve inversion, using more photometric observations than just the Lowell and GAIA DR2 datasets used to build the DAMIT model. The scale was obtained from fittings of their dimensionless model to all the occultation chords of all the events displayed in Figure \ref{fig:all_events} following \cite{durech2011}, resulting on average better matches to the chords than those displayed in Figure \ref{fig:all_events}. Their scaled 3D model is expected to replace the current 3D dimensionless DAMIT model for Diomedes. While their 3D dimensionless model represents an improvement over the current DAMIT model due to the use of updated photometric databases, some drawbacks are present in their use of occultation data in comparison with our procedures. We use fine tuned photometric tools to generate high quality occultation light curves with PRAIA (\cite{assafin2023}). And we determine ingress/egress instants with rigorous box modeling, taking into account Fresnel diffraction, star diameter, shadow velocity, exposure and dead time using SORA (see Section \ref{sec:instants}). \cite{hanus} do not perform photometry on occultation datasets and utilize the published Occult software provisional ingress/egress instants reported by the observers, which sometimes come from visual observations as in the case of the 1997 event. Moreover, while respecting the parameter uncertainties involved in the dimensionless model, our procedure allows for refining the pole orientation and rotation period using the more precise occultation data, which is not the case of \cite{hanus}. Thus, our procedures enable the improvement of DAMIT or any other input 3D dimensionless or not model. In a broader sense, we are further developing the method to allow for multi-occultation fittings, even when only one chord is present per occultation. Usually overlooked, multiple (three or more) mono-chord events occur for about 25\% of Trojans observed in the Lucky Star occultation campaigns. The use of generative methods incorporating observed and synthetic solar phase and rotation light curves in the procedures are also being implemented. In a future work, for Diomedes and other Trojans, a generalized procedure based on this pilot work will be presented, with results based on all available photometric and occultation datasets, as in \cite{hanus}.

This work presented pilot results in the context of a long term Lucky Star program aiming at establishing and improving the physical characteristics of Trojans, for better understanding this population and to contribute for the study of the formation and evolution of minor bodies in the Solar System.

%The conclusion text goes here.\vskip6pt
\vskip6pt
\enlargethispage{20pt}

%\ethics{Insert ethics statement here if applicable.}

%\dataccess{Insert details of how to access any supporting data here.}

\dataccess{Original occultation recording data: Zenodo doi:10.5281/zenodo.14166210 \cite{recordings}. Extracted occultation light curves: Zenodo doi:10.5281/zenodo.14153583
\cite{lightcurves}.}

\ack{
We thank the two anonymous referees for their comments which improved the text. H.D. is thankful for the support of the Coordenação de Aperfeiçoamento de Pessoal de Nível Superior - Brasil (CAPES) – Finance Code 001. The authors acknowledge the respective CNPq grants: B.E.M. 150612/2020-6; F.B.R. 316604/2023-2; M.A. 427700/2018-3, 310683/2017-3, 473002/2013-2; J.I.B.C. acknowledges grants 305917/2019-6, 306691/2022-1 (CNPq) and 201.681/2019 (FAPERJ). This work was carried out within the “Lucky Star” team that bring together the efforts of the Paris, Granada, and Rio groups, funded by the European Research Council under the European Community’s H2020 (ERC Grant Agreement No. 669416). Part of this work was supported by the Spanish projects PID2020-112789GB-I00 from AEI and Proyecto de Excelencia de la Junta de Andalucía PY20-01309. Financial support from the grant CEX2021-001131-S funded by MCIN/AEI/ 10.13039/501100011033 is also acknowledged. 
}

%\disclaimer{Insert disclaimer text here if applicable.}

%%%%%%%%%% Insert bibliography here %%%%%%%%%%%%%%


\begin{thebibliography}{9}

\bibitem[1]{bottke} Bottke, W. F. (Ed.). (2002). Asteroids III. University of Arizona Press.

\bibitem[2]{faure} Faure, G., Mensing, T. M., Faure, G., \& Mensing, T. M. (2007). Mars: The little planet that could. Introduction to Planetary Science: The Geological Perspective, 211-259.

\bibitem[3]{nesvorny} Nesvorny, D., Vokrouhlicky, D. and Morbidelli, A., 2013, Capture of trojans by jumping Jupiter. The Astrophysical Journal, 768(1):45.

\bibitem[4]{damit} Durech J., Sidorin V., Kaasalainen M., 2010, A\&A, 513, A46. doi:10.1051/0004-6361/200912693.

\bibitem[5]{wells1981} Wells D. C., Greisen E. W., Harten R. H., 1981, A\&AS, 44, 363.

\bibitem[6]{guga2016} Benedetti-Rossi G., Sicardy B., Buie M. W., et al., 2016, AJ, 152, 156.

\bibitem[7]{barry2015} Barry M. A. T., Gault D., Pavlov H., et al., 2015, Publications of the Astronomical Society of Australia, 32, e031.

\bibitem[8]{assafin2023} Assafin, M. (2023). Differential aperture photometry and digital coronagraphy with PRAIA. Planetary and Space Science, 239, 105816.

\bibitem[9]{gaia2022} Gaia Collaboration, Vallenari A., Brown A.~G.~A., Prusti T., de Bruijne J.~H.~J., Arenou F., Babusiaux C., et al., 2023, A\&A, 674, A1. doi:10.1051/0004-6361/202243940.

\bibitem[10]{sora} Gomes-J\'unior A. R., Morgado B. E., Benedetti-Rossi G., et al., 2022, MNRAS, 511, 1167.

\bibitem[11]{cantat2021} Cantat-Gaudin T., Brandt T., 2021, A\&A, 649, A124.

\bibitem[12]{nomad} Zacharias N., Monet D.~G., Levine S.~E., Urban S.~E., Gaume R., Wycoff G.~L., 2005, yCat, 1297. I/297.

\bibitem[13]{kervella2004} Kervella P., Th\'evenin F., Di Folco E., S\'egransan D., 2004, A\&A, 426, 297.

\bibitem[14]{mainzer2019} Mainzer A.~K., Bauer J.~M., Cutri R.~M., Grav T., Kramer E.~A., Masiero J.~R., Sonnett S., et al., 2019, PDSS, 251.

\bibitem[15]{durech2019} Durech J., Hanu{\v{s}} J., Van{\v{c}}o R., 2019, A\&A, 631, A2. doi:10.1051/0004-6361/201936341. 

\bibitem[16]{Belle1999} van Belle G. T., 1999, PASP, 111, 1515.

\bibitem[17]{archinal} Archinal, B.A., Acton, C.H., A’Hearn, M.F. et al., 2018, Report of the IAU Working Group on Cartographic Coordinates and Rotational Elements: 2015. Celest Mech Dyn Astr 130, 22. https://doi.org/10.1007/s10569-017-9805-5.

\bibitem[18]{kaasalainen} Kaasalainen M., Torppa J., 2001, Icar, 153, 24.

\bibitem[19]{neese} Neese, C., Ed. (2017). Asteroid Taxonomy V1.0. urn:nasa:pds:ast\textunderscore taxonomy::1.0. NASA Planetary Data System; https://doi.org/10.26033/e1p3-xm59.

\bibitem[20]{Krasinsky} Krasinsky G.A., Pitjeva E.V. , Vasilyev M.V. , Yagudina E.I., 2002, Icar, 158, 98.

\bibitem[21]{sicardy2011} Sicardy, B., Ortiz, J., Assafin, M. et al., 2011, A Pluto-like radius and a high albedo for the dwarf planet Eris from an occultation. Nature 478, 493–496.

\bibitem[22]{carry} Carry B., Peloton J., Le Montagner R., Mahlke M., Berthier J., 2024, A\&A, 687, A38.

\bibitem[23]{stephens} Stephens R.~D., 2009, MPBu, 36, 59.

\bibitem[24]{hanus} Hanu{\v{s}} J., Vokrouhlick{\'y} D., Nesvorn{\'y} D., {\v{D}}urech J., Stephens R., Benishek V., Oey J., et al., 2023, A\&A, 679, A56. doi:10.1051/0004-6361/202346022.

\bibitem[25]{sato} Sato I., Sarounova L., Fukushima H., 2000, Icar, 145, 25.

\bibitem[26]{usui} Usui F., Kuroda D., M{\"u}ller T.~G., Hasegawa S., Ishiguro M., Ootsubo T., Ishihara D., et al., 2011, PASJ, 63, 1117. doi:10.1093/pasj/63.5.1117.


\bibitem[27]{iras} Tedesco E.F., Noah P.V., Noah M., Price S.D., 2002, AJ, 123, 1056.

\bibitem[28]{durech2011} Durech J., Kaasalainen M., Herald D., et al., 2011, Icar, 214, 652.


\bibitem[29]{recordings} Dutra H., Assafin M., Sicardy B., et al., 2024. Video Recordings from a Stellar Occultation by Jupiter's Trojan Diomedes on November 1st, 2020. In
Philosophical Transactions of the Royal Society A. Zenodo. https://doi.org/10.5281/zenodo.14166210


\bibitem[30]{lightcurves} Dutra H., Assafin M., Sicardy B., et al., 2024. Light Curves from a Tri-chord Stellar Occultation by Jupiter's Trojan Diomedes in 2020 [Data set]. In Philosophical Transactions of the Royal Society A. Zenodo. https://doi.org/10.5281/zenodo.14153583


%%%%%%%%%%%%%%%%%%%%%%%%%%%%%%%%%%%%%%%%%%%%%%%%%%%%%%%%%%%%%%%%%%%%%%%%%%%%%%%%%%%

\end{thebibliography}
\end{document}